\documentstyle[12pt]{article}

\makeatletter
\def\eqnarray{\stepcounter{equation}\let\@currentlabel=\theequation
\global\@eqnswtrue
\global\@eqcnt\z@\tabskip\@centering\let\\=\@eqncr
$$\halign to \displaywidth\bgroup\@eqnsel\hskip\@centering
  $\displaystyle\tabskip\z@{##}$&\global\@eqcnt\@ne
  \hfil$\displaystyle{\hbox{}##\hbox{}}$\hfil
  &\global\@eqcnt\tw@ $\displaystyle\tabskip\z@
  {##}$\hfil\tabskip\@centering&\llap{##}\tabskip\z@\cr}
\def\@sect#1#2#3#4#5#6[#7]#8{\ifnum #2>\c@secnumdepth
    \def\@svsec{}\else
    \refstepcounter{#1}\edef\@svsec{\csname the#1\endcsname.\hskip 1em }\fi
    \@tempskipa #5\relax
    \ifdim \@tempskipa>\z@
    \begingroup #6\relax
    \@hangfrom{\hskip #3\relax\@svsec}{\interlinepenalty \@M #8\par}
    \endgroup
    \csname #1mark\endcsname{#7}\addcontentsline
    {toc}{#1}{\ifnum #2>\c@secnumdepth \else
     \protect\numberline{\csname the#1\endcsname}\fi
           #7}\else
    \def\@svsechd{#6\hskip #3\@svsec #8\csname #1mark\endcsname
          {#7}\addcontentsline
          {toc}{#1}{\ifnum #2>\c@secnumdepth \else
     \protect\numberline{\csname the#1\endcsname}\fi
           #7}}\fi
     \@xsect{#5}}
\def\label#1{\@bsphack\if@filesw {\let\thepage\relax
   \xdef\@gtempa{\write\@auxout{\string
   \newlabel{#1}{{\thesection.\@currentlabel}{\thepage}}}}}\@gtempa
   \if@nobreak \ifvmode\nobreak\fi\fi\fi\@esphack}
\def\@eqnnum{(\thesection.\theequation)}
\def\section{\setcounter{equation}{0} \@startsection {section}{1}{\z@}{-3.5ex
   plus -1ex minus -.2ex}{2.3ex plus .2ex}{\Large\bf}}
\newcount\@minsofar
\newcount\@min
\newcount\@cite@temp
\def\appendixname{Appendix}
\def\appendix{\par
  \def\pre@section{\appendixname{}}
  \setcounter{section}{1}
  \@addtoreset{equation}{section}
  \def\thesection{\Alph{section}}
  \def\theequation{\arabic{equation}}}

\def\appendix{\par
  \def\pre@section{\appendixname{}}
  \setcounter{section}{1}
  \@addtoreset{equation}{section}
  \def\thesection{\Alph{section}}
  \def\theequation{\arabic{equation}}}

\makeatother

\def\({\left(}
\def\){\right)}
\def\[{\left[}
\def\]{\right]}
\def\non{ \nonumber }

\def\debut{\begin{align}}
\def\fin{\end{align}}
\def\half{\textstyle{\frac 1 2}}

\def\l{\lambda}
\def\e{\epsilon}

\def\b{\beta}

\def\g{\gamma}
\def\a{\alpha}
\def\s{\sigma}

\def\l{\lambda}

\def\e{\epsilon}

\def\ds{\displaystyle}
\def\be{\begin{equation}}
\def\ee{\end{equation}}
\def\beq{\begin{eqnarray}}
\def\eeq{\end{eqnarray}}
\def\ov{\overline}

\begin{document}

\phantom{a}

\vspace{0.5cm}

\begin{center}
{\bf Emptiness Formation Probability and Quantum Knizhnik-Zamolodchikov 
Equation}
\end{center}
\phantom{a}

\vspace{1.5cm}

\centerline{H. E. Boos 
\footnote{on leave of absence from the Institute for High Energy Physics,
Protvino, 142284, Russia}}
\centerline{\it Max-Planck Institut f{\"u}r Mathematik}
\centerline{\it Vivatsgasse 7, 53111 Bonn, Germany} 

\phantom{a}

\vspace{0.5cm}

\phantom{a}

\centerline{V. E.  Korepin }
\centerline{\it C.N.~Yang Institute for Theoretical Physics}
\centerline{\it State University of New York at Stony Brook}
\centerline{\it Stony Brook, NY 11794--3840, USA}

\phantom{a}

\vspace{0.5cm}

\phantom{a}

\centerline{F.A. Smirnov
\footnote{Membre du CNRS}
}
\centerline{\it LPTHE, Tour 16, 1-er {\'e}tage, 4, pl. Jussieu}
\centerline{\it 75252, Paris Cedex 05, France}

\vspace{1cm}

\vskip2em
\begin{abstract}
\noindent
We consider the one-dimensional 
XXX spin 1/2 Heisenberg antiferromagnet at zero temperature
and zero magnetic field. We are interested in a probability  of formation of
a ferromagnetic string $P(n)$  in the antiferromagnetic ground-state. 
We call it emptiness formation probability [EFP]. We suggest  a new technique for computation of the EFP
in the inhomogeneous case. It is based on the quantum Knizhnik-Zamolodchikov 
equation [qKZ]. We calculate EFP for $n\le 6$
for inhomogeneous case. The homogeneous limit confirms our hypothesis about the relation
of  quantum correlations  and number theory.
We also make a conjecture  about a  structure of EFP for arbitrary $n$.
\end{abstract}

\newpage

\section{Introduction}

The Hamiltonian of the XXX Heisenberg spin chain can be written like this
\be
H = \sum_{i=-N}^N \>
(\s^x_i\s^x_{i+1}\; + \;\s^y_i\s^y_{i+1}\; +\; \s^z_i\s^z_{i+1}\;-1\;)
\label{H}
\ee
Here $2 N +1 $ is the length of the lattice and  $\s^x_i,\s^y_i,\s^z_i$ are
Pauli  matrices. We consider thermodynamic limit [ $N$ goes to infinity] .
The sign in front of the Hamiltonian indicates that we
are dealing with the antiferromagnetic case. We also imply
periodic boundary conditions.
The  model was solved by Bethe in 1931, see \cite{B}.
The ground state was  constructed by by Hulth\'{e}n, see \cite{H}.
We shall denote the ground state in the thermodynamic limit by ${| {\rm GS} \rangle}$ .
The emptiness formation probability (EFP) for the XXX model is
defined as follows:
\begin{equation}
P(n) = \langle {\rm GS} | \prod^n_{j=1} P_j | {\rm GS} \rangle,
\end{equation}
here ${P_j = S^z_j+\half}$ is the projector on the state 
with the spin up in the ${j}$-th lattice site. 
The integer $n$ has a meaning of a length of a continuous ferromagnetic string.
$P(n)$ is a probability that this string can appear 
in the antiferromagnetic  ground-state. The importance of EFP was emphasized  in \cite{KBI}.

 $P(n)$  can be represented as $n$-multiple integral.
The  integral representation follows from the work of  RIMS group.
RIMS approach is based on vertex operators and  bosonic representation of infinite-dimensional 
quantum algebras, see  \cite{JMN,JM}. The explicit formula for  $P(n)$ in XXX limit
was obtained in  \cite{KIEU}:

\be
P(n)=
\prod_{j=1}^n\int_{C}{d\lambda_j\;\over 2\pi i }\;
U_n(\l_1,\ldots,\l_n)\;T_n(\l_1,\ldots,\l_n)
\label{int}
\ee
where
\be
U_n(\l_1,\ldots,\l_n)\;=\;\pi^{{n(n+1)\over 2}}\>
{\prod_{1\leq k < j\leq n}\sinh{\pi(\lambda_j-\lambda_k)}
\over
\prod_{j=1}^n\sinh^n{\pi\lambda_j}
}
\label{U_n}
\ee
and
\be
T_n(\l_1,\ldots,\l_n)\;=\;
{\prod_{j=1}^n\l_j^{j-1}(\lambda_j+i)^{n-j}\over
\prod_{1\leq k < j\leq n}(\lambda_j-\lambda_k-i)
}
\label{T_n}
\ee

The  contour $C$  goes parallel to the real axis with the imaginary part confined  between 
$0$ and $-i$ for each integral.

In papers \cite{BK1,BK2,BKNS} we
 evaluated the integrals
for $n\le 5$. 
We  discovered that these  $P(n)$ can be expressed in terms of 
the values of  Riemann zeta function at odd arguments,  $\log 2$ and  rational coefficients.
We  conjectured that this is a general property 
for all  $P(n)$. In this paper we proved the property for $P(6)$.
We think that all  correlation functions 
\be
\langle {\rm GS} | \s^z_{i_1}\;\s^z_{i_2}\;\ldots \s^z_{i_m}| {\rm GS} \rangle
\label{cor}
\ee
also have this property.
Asymptotic behavior  of  $P(n)$ for large $n$ was studied in the papers
\cite{BK1},\cite{BKNS}, \cite{AK} and \cite{LNKS}.
The technique of calculation of these integrals, described  in the paper \cite{BK2} worked for 
$n=2,3,4$.
In the paper  \cite{BKNS} we  calculated $P(5)$ by means of  this technique.
However, these computations are so complicated
that it is problematic to generalize them to  the case $n=6$.
So  we start looking  for indirect methods of
 evaluation of the integrals (\ref{int}). 

It appeared to be useful to consider inhomogeneous case.
In this case there are more free parameters. We  call them  inhomogeneity parameters and denote by
$z_1,\ldots,z_n$.  The EFP in the inhomogeneous case  we shall
denote by $P_n(z_1,\ldots,z_n)$.
 Let us remind to the reader that 
inhomogeneous models were used for evaluation of correlation functions from the very  beginning.
For the massive regime of the XXZ model
the vertex operator approach  was developed in \cite{JMN,JM}. It  allowed to
express the correlation functions in terms of the trace functions. Special
combinations of these trace functions satisfy the  
quantum Knizhnik-Zamolodchikov equation (qKZ), see \cite{KZ,FR}. 
Later in paper \cite{JM1} Miwa and Jimbo suggested that the correlation
functions in the gap-less regime directly satisfy the qKZ equation. 
Since the XXX model belongs to the gap-less regime we shall
use  qKZ for evaluation of  EFP in the inhomogeneous
case. We suggest a general ansatz for  $P_n(z_1,\ldots,z_n)$, see (3.20).
This constitute a new method  for computation of the EFP. 
On the other hand, it is easy to generalize the technique
explained in \cite{BK2} to the inhomogeneous case and calculate the EFP directly
[for short ferromagnetic strings]. When we can compare results, they coincide.

The paper is organized as follows. 
In the Section 2 we discuss the relation of EFP to the qKZ
and derive three important properties of the EFP in the inhomogeneous case.
In Section 3 we apply a generalization of the technique described in \cite{BK2} to
the inhomogeneous case and compute  $P_n$ directly for $n\le 4$.
Then we check that these $P_n$  satisfy all the properties, which follow from qKZ. This helps us to formulate
a general {\bf ansatz} for $P_n$ in the inhomogeneous case.
Further, we suggest a new way of computing  $P_n$. One can  use the 
ansatz and general properties of $P_n$ [which follow from the qKZ].
In this way we get the explicit expressions for $P_5$ and $P_6$ in the inhomogeneous
case. 
In Section 4 we discuss the homogeneous limit of $P(n)$ for $n\le 6$.
In particular, when $n\le 5$ we reproduce our previous results, obtained
in \cite{BK1,BK2,BKNS}.
We also get the analytic expression  for $P(6)$ in the homogeneous limit. 
Having this answer we can
compare it with the numerical value for $P(6)$ obtained by the DMRG 
method in \cite{BKNS}.
We also discuss the structure
of EFP in the homogeneous case and offer some plausible conjectures.
In the last Section 6 we discuss the results and outline some possible ways
of a further progress.

\section{The EFP in the inhomogeneous case and the qKZ equation.}

We believe that consideration of the inhomogeneous case instead of the
homogeneous one can give us a new information about the EFP and other 
correlation functions.

A method of calculation of correlation functions, which we use  was found in the papers \cite{JMN, JM}.
 It  is based on theory of infinite-dimensional quantized algebras and vertex operators. 
We shall need elements of this method.
Let us introduce some notations.

We use 
the R-matrix:
\be
R(\l)=
\frac { R_0(\l)} {\l+\pi i}
\left(
\begin{array}{cccc}
\l+\pi i&0&0&0\\
0&-\l&\pi i&0\\
0&\pi i&-\l&0\\
0&0&0&\l+\pi i
\end{array}
\right)
\label{R-m}
\ee
where
$$
R_0(\l)=-\frac
{\Gamma\(\frac \l {2\pi i}\)\Gamma\(\frac 1 2-\frac \l {2\pi i}\)}
{\Gamma\(-\frac \l {2\pi i}\)\Gamma\(\frac 1 2+\frac \l {2\pi i}\)},
$$
notice that
$$R_0(\l)R_0(-\l)=1$$
This R-matrix appears in the rational limit from XXZ R-matrix, it is related 
to usual XXX R-matrix by
$$R(\l)=(\sigma ^z\otimes I )R_{XXX}(\l)(I\otimes \sigma ^z)$$
(similar transformation is needed when 
obtaining form factors of $SU(2)$-invariant Thirring
model from SG ones \cite{book}).

This R-matrix (\ref{R-m}) satisfies the equation:
\be
R(-\pi i)=
\left(
\begin{array}{cccc}
0&0&0&0\\
0&1&1&0\\
0&1&1&0\\
0&0&0&0
\end{array}
\right)
\label{proj}
\ee

Following \cite{JM} we introduce functions $g_n$ which satisfy
the qKZ equations on level $-4$ \cite{FR}. 
We write the qKZ equations in their original form \cite{book,sm} which
takes into account explicitly symmetry:
\beq
&g_n(\l _1,\cdots ,\l _{j+1},\l _j,\cdots,\l _{2n})_
{\e _1,\cdots ,\e _{j+1}',\e _j',\cdots,\e _{2n}}=&\label{symm}\\
&=R(\l _j -\l _{j+1})^{\e _{j},\e _{j+1}}_{\e _{j}',\e _{j+1}'}
\  \ g_n(\l _1,\cdots ,\l _{j},\l _{j+1},\cdots,\l _{2n})
_{\e _1,\cdots ,\e _{j},\e _{j+1},\cdots,\e _{2n}}
\nonumber
\eeq
\be
g_n(\l _1,\cdots ,\l _{2n-1},\l _{2n}+2\pi i)_
{\e _1,\cdots ,\e _{2n-1},\e _{2n}}
=
g_n(\l _{2n},\l _1,\cdots ,\l _{2n-1})
_{\e _{2n},\e _1,\cdots ,\e _{2n-1}}
\label{Rie}
\ee

Solutions to these equations are meromorphic functions with possible
singularities at the points 
$$ \Im (\l _j-\l_k)=\pi  l, \quad l\in {\mbox{\bf Z}}\backslash 0$$
For application to the correlation functions we are interested in a
particular solution  $g_n$ described in details in \cite{JMN,JM}. 
Detailed study of
this solution will be performed in a future publication, in the present
paper we need only limited information about it.
First, the $g_n$ is regular at $ \Im (\l _j-\l_k)=\pm\pi $. 

Moreover, much can be said
about its values at these points \cite{JM} :
\beq
&g_n(\l _1,\cdots ,\l _{j-1},\l _{j},\l _j-\pi i,\l _{j+2},\cdots,\l _{2n})_
{\e _1,\cdots ,\e _{j-1},\e _{j},\e _{j+1},\e _{j+2},\cdots,\e _{2n}}=&
\nonumber\\
&=\delta_{\e _{j},-\e _{j+1}}
\ g_{n-1}(\l _1,\cdots ,\l _{j-1},\l _{j+2},\cdots,\l _{2n})
_{\e _1,\cdots,\e _{j-1},\e _{j+2}, \cdots,\e _{2n}}
&\label{val1}
\eeq
Together with the symmetry (\ref{symm}), and (\ref{proj}) this equation implies
\beq
&\sum\limits_{\e _j=-\e _{j+1}}g_n(\l _1,\cdots ,\l _{j-1},
\l _{j},\l _j+\pi i,\l _{j+2},\cdots,\l _{2n})_
{\e _1,\cdots ,\e _{j-1},\e _{j},\e _{j+1},\e _{j+2},\cdots,\e _{2n}}
=&\nonumber\\
&=
g_{n-1}(\l _1,\cdots ,\l _{j-1},\l _{j+2},\cdots,\l _{2n})
_{\e _1,\cdots,\e _{j-1},\e _{j+2}, \cdots,\e _{2n}}
&\label{val2}
\eeq

The emptiness formation probability $P_n$ is related to $g_n$ as follows
\be
P_n(z_1,\cdots ,z_n)=
g_n\(\pi z_1,\cdots, \pi z_n,\pi (z_n+i),\cdots,\pi (z_1+i)\)
_{-,\cdots ,-,+,\cdots ,+}
\label{Png}
\ee

Now we want to establish some general properties of $P_n$ following
from (\ref{symm}), (\ref{Rie}), (\ref{val1}), (\ref{val2}).

\noindent
1. Symmetry. {\it The function $P_n(z_1,\cdots ,z_n)$ is symmetric.}

\noindent{\it Proof.} Obviously, it is enough to show that
$$P_n(\cdots,z_j,z_{j+1},\cdots)=P_n(\cdots,z_{j+1},z_j,\cdots)$$
This identity follows from (\ref{symm}) and from the fact that the R-matrix
acts diagonally on the indices $-,-$ and $+,+$:
\beq
&P_n(\cdots,z_{j},z_{j+1},\cdots)=&\nonumber\\
&=
g_n\(\cdots,\pi z_{j},\pi z_{j+1},\cdots,\pi (z_{j+1}+i),\pi(z_j+i)\cdots\)
_{\cdots, -,-,\cdots ,+,+,\cdots}=&\nonumber\\
&=
R_0(\pi (z_{j+1}-z_j))R_0(\pi (z_{j}-z_{j+1}))&\nonumber\\
&\times
g_n\(\cdots,\pi z_{j+1},\pi z_{j},\cdots,\pi (z_{j}+i),\pi(z_{j+1}+i)\cdots\)
_{\cdots, -,-,\cdots ,+,+,\cdots}=\non\\&=P_n(\cdots,z_{j+1},z_{j},\cdots)
&\nonumber
\eeq
\hfill{\bf QED}

\noindent
2. Vanishing. {\it The function $P_n(z_1,\cdots ,z_n)$ 
vanishes when $z_k=z_j+i$.}

\noindent{\it Proof.}  
Due to the previous property it is sufficient to consider the case
$k=1$, $j=2$. Let us put, first, $z_1=z+i$, $z_{2}=z'$, 
then we shall take the limit $z\to z'$.
\beq
&P_n(z+i,z',\cdots ,z_n)=&\nonumber\\
&=
g_n\(\pi (z+i),\pi z',\cdots, \pi z_n,\pi (z_n+i),\cdots,\pi(z'+i),\pi (z+2i)\)
_{-,-,\cdots, -,+,\cdots ,+,+}=&\nonumber\\
&=
g_n\(\pi z,\pi (z+i),\pi z',\cdots, \pi z_n,\pi (z_n+i),\cdots,\pi(z'+i),\)
_{+,-,-,\cdots, -,+,\cdots ,+}&\nonumber
\eeq
where we used (\ref{Rie}). Consider the limit $z\to z'$. As it has been 
explained singularities
do not occur for $\Im (\l _j -\l _k)=\pm \pi$. Moreover, the final result 
contains the fragment
$$g_n(\cdots, \pi (z+i),\pi z\cdots)_{\cdots ,-,-,\cdots }$$
which implies that the result vanishes due to (\ref{val1}). Because of absence 
of singularities
this zero does not interfere with any pole, so
\be
P_n(z+i,z,\cdots ,z_n)=0
\label{vanish}
\ee
\hfill{\bf QED}

\noindent
3. Normalization. 
{\it The following asymptotic holds for $z_1\to\infty$ along the real axis:}
\be
P_n(z_1,z_2,\cdots,z_n)\to \half P_{n-1}(z_2,\cdots,z_{n})\label{norm}
\ee

\noindent{\it Proof.} 
One more property of the solution $g_n$ will be important for us. Using
the integral formula from \cite{JM} one can show that
$g_n(\l_1,\cdots ,\l _{2n-2},\l,\l+\pi i)$ behaves as 
$O(1)$ when $\l\to\infty +i\kappa$ 
where $\kappa$ is a finite number. 
The leading term of asymptotic does not depend on
$\kappa$. We shall use notation:
\be 
g_n(\l_1,\cdots ,\l _{2n-2},\l,\l+\pi i)_
{\e_1,\cdots ,\e_{2n-2},\e_{2n-1},\e_{2n}}\to 
\widehat{g}_{n}(\l_1,\cdots ,\l _{2n-2})_{\e_1,\cdots ,\e_{2n-2}\ ;\ 
\e_{2n-1},\e_{2n}}
\label{hat}
\ee
The function $\widehat{g}_n$ possesses important property of symmetry
with respect to last two indices because
\beq
&g_n(\l_1,\cdots ,\l _{2n-2},\l,\l+\pi i)_{\e_1,\cdots ,\e_{2n-2},\e_{2n-1},\e_{2n}}=&
\nonumber\\
&=
 g_n(\l-\pi i,\l_1,\cdots ,\l _{2n-2},\l)_{\e_{2n}\e_1,\cdots ,\e_{2n-2},\e_{2n-1}}&
\nonumber\\
&=
R(\l_1-\l+\pi i) _{\e_1,\e_{2n}}^{\e_1',\sigma _{2n-3}}
\cdots 
R(\l_{2n-2}-\l+\pi i)_{\e_{2n-2},\sigma_1}^{e_{2n-2}',\e _{2n}'}&\nonumber\\
&\times
g_n(\l_1,\cdots ,\l _{2n-2},\l-\pi i,\l)_{\e_1',\cdots ,
\e_{2n-2}',\e_{2n}',\e_{2n-1}}\rightarrow &\nonumber\\
&
\rightarrow
\mbox{sign}(\e_{2n-1},\e_{2n})\,
\widehat{g}_{n}(\l_1,\cdots ,\l _{2n-2})_{\e_1,\cdots ,\e_{2n-2}\ ;\ \e_{2n},\e_{2n-1}}&
\label{ghat}
\eeq
with the sign function
\be
\mbox{sign}(\e_1,\e_2)\;=\;
\cases{
{\ds -1 } & if ${\ds \e_1=\e_2}$ \cr
{\ds \>1 } & if ${\ds \e_1=-\e_2}$ \cr
}
\ee
where we used the asymptotic of the R-matrix
$$R(\l)\ \longrightarrow {\hskip -.8cm}_{_{\l\to\infty}}\ (-i)\cdot\mbox{diag}(1,-1,-1,1)$$
From the equation (\ref{ghat}) we conclude that
$$
\widehat{g}_{n}(\l_1,\cdots ,\l _{2n-2})_{\e_1,\cdots ,\e_{2n-2}\ ;\ \e_{2n-1},\e_{2n}}=0
\quad \mbox{if} \quad \e_{2n-1}=\e_{2n}
$$
\be
\widehat{g}_{n}(\l_1,\cdots ,\l _{2n-2})_{\e_1,\cdots ,\e_{2n-2}\ ;\ +,-}=
\widehat{g}_{n}(\l_1,\cdots ,\l _{2n-2})_{\e_1,\cdots ,\e_{2n-2}\ ;\ -,+}
\label{ghat+-}
\ee

The relation (\ref{ghat+-}) allows to calculate 
$\widehat{g}_{n}(\l_1,\cdots ,\l _{2n-2})_{\e_1,\cdots ,\e_{2n-2}\ ;\ +,-}$:
\beq
&g_{n-1}(\l_1,\cdots ,\l _{2n-2})_{\e_1,\cdots ,\e_{2n-2}}=&\nonumber\\
&=
\widehat{g}_{n}(\l_1,\cdots ,\l _{2n-2})_{\e_1,\cdots ,\e_{2n-2}\ ;\ +,-}+
\widehat{g}_{n}(\l_1,\cdots ,\l _{2n-2})_{\e_1,\cdots ,\e_{2n-2}\ ;\ -,+}=&
\nonumber\\
&=
2\ \widehat{g}_{n}(\l_1,\cdots ,\l _{2n-2})_{\e_1,\cdots ,\e_{2n-2}\ ;\ +,-}&
\nonumber
\eeq
Now the normalization (\ref{norm}) follows from
\beq
&P_n(z_1,z_2,\cdots ,z_n)=&\nonumber\\
&=
g_n\(\pi z_1,\pi z_2,\cdots, \pi z_n,\pi (z_n+i),\cdots,\pi (z_2+i),\pi (z_1+i)\)
_{-,-,\cdots, -,+,\cdots ,+,+}=&\nonumber\\
&=
g_n\(\pi z_2,\cdots, \pi z_n,\pi (z_n+i),\cdots,\pi (z_2+i),\pi (z_1+i),\pi (z_1+2i)\)
_{-,\cdots, -,+,\cdots ,+,+,-}&\nonumber\\
&
\longrightarrow\hskip -.8cm _{_{z_1\to\infty}}
\ \widehat{g}_n\(\pi z_2,\cdots, \pi z_n,\pi (z_n+i),\cdots,\pi (z_2+i)\)
_{-,\cdots, -,+,\cdots ,+\ ;\ +,-}=&\nonumber\\
&=
\half \ P_{n-1}(z_2,\cdots ,z_n)&
\eeq
where the formula (\ref{hat}) was used. 
\hfill{\bf QED}

The above three properties will be very useful for further consideration.

\section{Explicit expressions for EFP.}

In paper \cite{JM1} Jimbo and Miwa suggested an integral representation 
as a solution to the qKZ (\ref{symm}) - (\ref{Rie}) which also satisfies the
property (\ref{val1}). Using the relation (\ref{Png})
one gets the integral representation
which is the direct generalization of the formula (\ref{int}) to the 
inhomogeneous case
$$
P_n(z_1,\ldots,z_n) = 
\pi^{{n(n+1)\over 2}}\prod_{k<j}{\sinh{(\pi(z_k-z_j))}\over\pi(z_k-z_j)}\;
\int_{-i/2-\infty}^{-i/2+\infty}{d\l_1\over 2\pi i}\ldots 
\int_{-i/2-\infty}^{-i/2+\infty}{d\l_n\over 2\pi i}
\prod_{k<j}{\sinh(\pi(\l_j-\l_k))\over (\l_j-\l_k-i)}\cdot
$$
\be
\cdot\,{\prod_{j=1}^n(\prod_{k=1}^{j-1}(\l_j-z_k-i)\prod_{k=j+1}^n(\l_j-z_k))\over
\prod_{j=1}^n\prod_{k=1}^n\sinh(\pi(\l_j-z_k))}
\label{intinhom}
\ee
where for the moment we will consider the inhomogeneity parameters $z_j$
as distinct real numbers. These parameters  also may be considered as arbitrary complex numbers,
but the integration contours in this case should be taken in such a way that
they separate the singularities of the integrand in the same manner
as the contours in the above formula (\ref{intinhom}) for the real values 
$z_j$.

For evaluation of the integral (\ref{intinhom}) we can use  
the technique described  in \cite{BK2}.  We obtain for the first four values of  $P_n(z_1,\ldots,z_n)$ :
\be
P_1 = {1\over 2}
\label{P1}
\ee
\be
P_2 = {1\over 3} + A_{2,1}(z_1,z_2)\;G(z_1-z_2)
\label{P2}
\ee
\be
P_3 = {1\over 4} + 
A_{3,1}(z_1,z_2|z_3)G(z_1-z_2)+A_{3,1}(z_1,z_3|z_2)G(z_1-z_3)+
A_{3,1}(z_2,z_3|z_1)G(z_2-z_3)
\label{P3}
\ee
$$
P_4 = {1\over 5} + 
A_{4,1}(z_1,z_2|z_3,z_4)G(z_1-z_2)+A_{4,1}(z_1,z_3|z_2,z_4)\;G(z_1-z_3)+
A_{4,1}(z_1,z_4|z_2,z_3)\;G(z_1-z_4)+
$$
$$
A_{4,1}(z_2,z_3|z_1,z_4)\;G(z_2-z_3)+
A_{4,1}(z_2,z_4|z_1,z_3)\;G(z_2-z_4)+A_{4,1}(z_3,z_4,|z_1,z_2)\;G(z_3-z_4)+
$$
$$
A_{4,2}(z_1,z_2|z_3,z_4)G(z_1-z_2)G(z_3-z_4)+
A_{4,2}(z_1,z_3|z_2,z_4)G(z_1-z_3)G(z_2-z_4)+
$$
\be
A_{4,2}(z_1,z_4|z_2,z_3)G(z_1-z_4)G(z_2-z_3)
\label{P4}
\ee
where
\be
G(x)=(x^2+1)(\b(ix)+\b(-ix))=2\sum_{k=1}^{\infty}(-1)^k\cdot k\cdot
{x^2+1\over x^2+k^2}
\label{G}
\ee
$$
\b(x)={1\over 2}(\psi({1+x\over2})-\psi({x\over2}))=
\sum_{k=0}^{\infty}{(-1)^k\over x+k}
$$
$$
\psi(x)={d\over dx}\ln{\Gamma(x)}
$$
and the functions
\beq
&\ds A_{2,1}(z_1,z_2)={\ov Q}_{2,1}(z_1,z_2) & \nonumber\\
&\ds A_{3,1}(z_1,z_2|z_3)={{\ov Q}_{3,1}(z_1,z_2|z_3)\over z_{13}z_{23}}
& \nonumber\\
&\ds A_{4,2}(z_1,z_2|z_3,z_4)\;=\;{{\ov Q}_{4,2}(z_1,z_2|z_3,z_4)\over 
z_{13}z_{14}z_{23}z_{24}} &\nonumber\\
&\ds A_{4,1}(z_1,z_2|z_3,z_4)\;=\;{{\ov Q}_{4,1}(z_1,z_2|z_3,z_4)\over 
z_{13}z_{14}z_{23}z_{24}} &\nonumber\\
\label{A}
\eeq
with the polynomials
\be
{\ov Q}_{2,1}(z_1,z_2)\;=\;{1\over 6}
\label{Q21}
\ee
\be
{\ov Q}_{3,1}(z_1,z_2|z_3)\;=\;{1\over 12}(\;z_{13}z_{23}-1)
\label{Q31}
\ee
\be
{\ov Q}_{4,2}(z_1,z_2|z_3,z_4)\;=\;
{1\over 36}\biggl\{ (z_{13}z_{24}-1)(z_{14}z_{23}-1)\;+\;
{2\over 5} (z_{12}^2\,+\,{3\over 2})(z_{34}^2\,+\,{3\over 2})\,+\,{3\over 2}
\biggr\}
\label{Q42}
\ee
\be
{\ov Q}_{4,1}(z_1,z_2|z_3,z_4)\;=\;
2\,{\ov Q}_{4,2}(z_1,z_2|z_3,z_4)-
{1\over 60}(z_{12}^2+4)(z_{34}^2+1)
\label{Q41}
\ee 
and by definition
$$
z_{jk} = z_j-z_k
$$

The function $G(x)$ (\ref{G}) is actually real for real $x$ and even 
\be
G(-x)=G(x)
\label{GG}
\ee
Besides, for small $x$ one has an expansion
\be
G(x) = -2(1+x^2)\sum_{k=0}^{\infty}(-1)^k x^{2k} \zeta_a(2k+1)
\label{Gexp}
\ee
with the alternating zeta series defined in (\ref{za}).
In particular,
\be
G(0)=-2\ln{2}
\label{G0}
\ee
The following limit is well defined
\be
G(\pm \infty)=-{1\over 2}
\label{Ginfty}
\ee
We shall also need the following relation
\be
G(x-i) = \a(x)\;+\;\gamma(x)\,G(x)
\label{Gx-i}
\ee
where 
\be
\a(x)\;=\;-{(x-2i)\over(x-i)},\quad
\gamma(x)\;=\;-{x(x-2i)\over(x+i)(x-i)}
\label{algamma}
\ee
In particular,
\be
G(\pm\,i)\;=\;-2
\label{Gpmi}
\ee

Looking at the answers (\ref{P1}-\ref{P4}) we can conclude that
they satisfy the general three properties from the previous section
and one more property, namely, {\it the translational invariance:}
\be
P_n(z_1+a,\ldots,z_n+a)\;=\;P_n(z_1,\ldots,z_n)
\label{transl}
\ee
which actually follows from the integral representation (\ref{intinhom}).
This property means that $P_n$ depends only on differences of $n$ parameters $z_j$.

One can also suggest {\it a general ansatz :}\\
\be
P_n(z_1,\ldots,z_n)\;=\;\sum_{l=0}^{[{n\over 2}]} 
\bigl\{
A_{n,l}(z_1,z_2|z_3,z_4|\ldots |z_{2l-1},z_{2l}|z_{2l+1}\ldots z_n)
\prod_{j=1}^lG(z_{2j-1}-z_{2j}) + \mbox{permutations}\bigr\}
\label{Pninhom}
\ee
where $A_{n,l}(z_1,z_2|z_3,z_4|\ldots |z_{2l-1},z_{2l}|z_{2l+1}\ldots z_n)$ 
are some rational functions 
which depend only on differences $z_j-z_k$. 
We also imply that these functions are symmetric under independent 
transpositions $z_1\leftrightarrow z_2, z_3\leftrightarrow z_4,\ldots,
z_{2l-1}\leftrightarrow z_{2l}$ and any permutation of the residual
variables $z_{2l+1}\ldots z_n$. Also we require a symmetry of this function 
under transposition of any pair $z_{2r-1}z_{2r}\leftrightarrow 
z_{2s-1}z_{2s}$
for $r,s\le l$. It is implied that the 
permutations in (\ref{Pninhom}) do not 
involve those sets of variables for which the
function $A_{n,l}$ is already symmetric. 
We also take  
\be
A_{n,0}(z_1,\ldots,z_n)={1\over n+1}
\label{An0}
\ee
and \footnote{Since $A_{n,0}$ are just constants below we will not 
write their arguments any more.}
\be
A_{n,l}(z_1,z_2|z_3,z_4|\ldots |z_{2l-1},z_{2l}|z_{2l+1}\ldots z_n)=
{Q_{n,l}(z_1,z_2|z_3,z_4|\ldots |z_{2l-1},z_{2l}|z_{2l+1}\ldots z_n)
\over \prod_{1\le j<k\le n}(z_j-z_k)}
\label{Anl}
\ee
where $Q_{n,l}$ are some polynomials of maximum 
power $n-1$ for each variable $z_k$.
Moreover we expect that these polynomials have factorized form, namely,
$$
Q_{n,l}(z_1,z_2|z_3,z_4|\ldots |z_{2l-1},z_{2l}|z_{2l+1}\ldots z_n)=
(z_1-z_2)(z_3-z_4)\ldots (z_{2l-1}-z_{2l})\prod_{2l+1\le j<k\le n}(z_j-z_k)
\cdot
$$
\be
{\ov{Q}}_{n,l}(z_1,z_2|z_3,z_4|\ldots |z_{2l-1},z_{2l}|z_{2l+1}\ldots z_n)
\label{Q}
\ee
where the polynomials 
${\ov{Q}}_{n,l}(z_1,z_2|z_3,z_4|\ldots |z_{2l-1},z_{2l}|z_{2l+1}\ldots z_n)$
depend only on differences of variables $z_j-z_k$. Also they
have the same symmetry properties as the function \\
$A_{n,l}(z_1,z_2|z_3,z_4|\ldots |z_{2l-1},z_{2l}|z_{2l+1}\ldots z_n)$ described
above. The maximum powers for each variable may be readily calculated
from the corresponding power of the polynomial $Q_{nl}$.

Let us rewrite the properties 2 and 3 from the Section 2 in the following form
\be
P_n(z_1,\ldots,z_{n-2},z_{n-1},z_{n-1}\pm i)\;=\; 0
\label{prop2}
\ee
\be
\lim_{z_n\rightarrow\infty}P_n(z_1,\ldots,z_{n-1},z_n)={1\over 2} 
P_{n-1}(z_1,\ldots,z_{n-1})
\label{rec}
\ee

As we shall see below together
with our main ansatz (\ref{Pninhom}) they completely fix the answer
for $P_5(z_1,\ldots,z_5)$ and $P_6(z_1,\ldots,z_6)$. Therefore we may
expect that it also happens for any $P_n(z_1,\ldots,z_n)$.

Let us start our consideration from the analysis of the relation (\ref{rec}).
First let us make a simple observation that it has an obvious solution
corresponding to the case when all the variables $z_j$ go to
infinity \footnote{The limits $z_n\rightarrow\infty,\ldots,
z_1\rightarrow\infty$ should be taken consequently}
i.e.
\be
P_n(\infty,\ldots,\infty)\;=\;{1\over 2^n}
\label{inftylimit}
\ee
 
Now we can analyze corollaries of the relation (\ref{rec}) and our
general ansatz (\ref{Pninhom}).
It can be seen that this is equivalent to the following
chain of recurrent relations for the functions $A_{n,l}$
$$
A_{2k+1,l}(z_1,z_2|z_3,z_4|\ldots |z_{2l-1},z_{2l}|
z_{2l+1}\ldots z_{2k},\infty)-
$$
$$
{1\over 2}\bigl\{
A_{2k+1,l+1}(z_1,z_2|z_3,z_4|\ldots |\infty,z_{2l+1}|z_{2l+2},\ldots,
z_{2k}) + \mbox{permutations}\bigr\}=
$$
\be
{1\over 2}
A_{2k,l}(z_1,z_2|z_3,z_4|\ldots |z_{2l+1},\ldots,z_{2k})\quad
\mbox{for} \quad l=0,1,\ldots,k
\label{rec1}
\ee

$$
A_{2k,l}(z_1,z_2|z_3,z_4|\ldots |z_{2l-1},z_{2l}|
z_{2l+1}\ldots z_{2k-1},\infty)-
$$
$$
{1\over 2}\bigl\{
A_{2k,l+1}(z_1,z_2|z_3,z_4|\ldots |\infty,z_{2l+1}|z_{2l+2},\ldots,
z_{2k-1}) + \mbox{permutations}\bigr\}=
$$
\be
{1\over 2}
A_{2k-1,l}(z_1,z_2|z_3,z_4|\ldots |z_{2l+1},\ldots,z_{2k-1})\quad
\mbox{for} \quad l=0,1,\ldots,k-1
\label{rec2}
\ee
where permutations 
are taken only for the variables $z_{2l+1},z_{2l+2},\ldots z_{2k}$ for
(\ref{rec1}) and $z_{2l+1},z_{2l+2},\ldots z_{2k-1}$ for (\ref{rec2})
in such a way that the variables within symmetric sets are not permuted.
In the formula (\ref{rec1}) it is also implied that
\be
A_{2k+1,k+1}\;=\;0.
\label{A2k+1k+1}
\ee

Up to the moment it is not completely clear how to write down
in a closed form all the equations for the functions $A_{n,k}$ which
follow from the relation (\ref{prop2}).

Let us write them down together with the equations (\ref{rec1}) and
(\ref{rec2}) consequently for $n=2,3,4,5,6$. 

$\quad$\\
{\it The case $n=2$}\\
Since $k=1$ in this case we have only one equation in the set (\ref{rec2})
\be
A_{2,0}\,-\,{1\over 2} A_{2,1}(\infty,z_1)\;=\;{1\over 2}A_{1,0}
\label{eq2inf}
\ee
which is obvious because as it can be seen from eqs. (\ref{P1}-\ref{P2}),
(\ref{A}) and (\ref{Q21}) all the functions $A$ here are
just numbers, namely,
$$
A_{1,0}\;=\;{1\over 2}
$$
$$
A_{2,0}\;=\;{1\over 3},\quad A_{2,1}(z_1,z_2)\;=\;{1\over 6}
$$

The second property, namely, the formula (\ref{prop2}) is also trivially
satisfied because using eq. (\ref{algamma}), (\ref{Gpmi}) 
we have from (\ref{P2})
$$
P_2(z_1,z_1+i)\;=\;A_{2,0}\,+\,A_{2,1}(z_1,z_1+i)G(-i)\;=\;
$$
\be
A_{2,0}\,-\,2\,A_{2,1}(z_1,z_1+i)\;=\;0
\label{eq2i}
\ee
So the equation corresponding to the second property is just the
last relation in formula (\ref{eq2i}).

$\quad$\\
{\it The case $n=3$}\\
Since $n$ is odd the first property (\ref{rec}) corresponds
to the relations (\ref{rec1}) with $k=1$. Here we have two 
of them
$$
A_{3,1}(z_1,z_2|\infty)\;=\;{1\over 2} A_{2,1}(z_1,z_2)
$$
\be
A_{3,0}(z_1,z_2|\infty)\,-\,
{1\over 2}\biggl\{A_{3,1}(\infty,z_1|z_2)\,+\,A_{3,1}(\infty,z_2|z_1)\biggr\}
\;=\;{1\over 2} A_{2,1}(z_1,z_2)
\label{eq3inf}
\ee

The second property (\ref{prop2}) i.e. the equation
\be
P_3(z_1,z_2,z_2+i)\;=\;0
\label{P3i}
\ee
is equivalent to the following two relations
$$
A_{3,1}(z_1,z_2|z_2+i)\,+\,\g(z_{12})A_{3,1}(z_1,z_2+i|z_2)\;=\;0
$$
\be
A_{3,0}\,+\,\a(z_{12})A_{3,1}(z_1,z_2+i|z_2)\,-\,2\,A_{3,1}(z_2,z_2+i|z_1)\;=\;0
\label{eq3i}
\ee
Substituting here $A_{3,1}$ defined in (\ref{A}) and (\ref{Q31}) 
($A_{3,0}=1/4$)
one can easily check that these two equations are satisfied.

$\quad$\\
{\it The case $n=4$}\\
As in the previous case the first property (\ref{rec}) is equivalent 
to the set of two relations (see (\ref{rec2}) for $k=2$)
$$
A_{4,1}(z_1,z_2|z_3,\infty)\,-\,{1\over 2} A_{4,2}(z_1,z_2|\infty,z_3)
\;=\;{1\over 2} A_{3,1}(z_1,z_2|z_3)
$$
\be
A_{4,0}\,-\,
{1\over 2}\biggl\{A_{4,1}(z_1,\infty|z_2,z_3)\,+\,A_{4,1}(z_2,\infty|z_1,z_3)
\,+\,A_{4,1}(z_3,\infty|z_1,z_2)\biggr\}
\;=\;{1\over 2} A_{3,0}
\label{eq4inf}
\ee

The Property 2 (\ref{prop2}) for $n=4$ is equivalent to the following
set of equations
$$
\g(z_{23})A_{4,2}(z_1,z_3|z_2,z_3+i)\,+\,\g(z_{13})A_{4,2}(z_1,z_3+i|z_2,z_3)
\;=\;0
$$
$$
A_{4,1}(z_1,z_2|z_3,z_3+i)\,-\,2\,A_{4,2}(z_1,z_2|z_3,z_3+i)
\;=\;0
$$
$$
A_{4,1}(z_1,z_3|z_2,z_3+i)\,+\,\g(z_{13})A_{4,1}(z_1,z_3+i|z_2,z_3)
\,+\,\a(z_{23})A_{4,2}(z_1,z_3|z_2,z_3+i)\;=\;0
$$
\be
A_{4,0}\,+\,\a(z_{13})A_{4,1}(z_1,z_3+i|z_2,z_3)\,+\,\a(z_{23})
A_{4,1}(z_2,z_3+i|z_1,z_3)\,-\,2\,A_{4,1}(z_3,z_3+i|z_1,z_2)\;=\;0
\label{eq4i}
\ee

Again using eqs. (\ref{A}) and (\ref{Q42}-\ref{Q41}) one can check
that the equations (\ref{eq4inf}-\ref{eq4i}) are satisfied.

So we have established that, in fact, the formulae (\ref{A}) with
the definitions (\ref{Q21}-\ref{Q41}) provide the solution
to the equations (\ref{eq2inf}) and (\ref{eq2i}) for $n=2$,
eqs. (\ref{eq3inf}) and (\ref{eq3i}) for $n=3$,
eqs. (\ref{eq4inf}) and (\ref{eq4i}) for $n=4$.

We may also go an opposite way, namely, demand all these equations, solve
them in framework of the ansatz (\ref{Pninhom}-\ref{Q}) 
and then come to the result for $P_n$.
Let us mention that this way seems to be much easier than the direct calculation
of the integral (\ref{intinhom}) as we did before.

For $n=5,6$ we would like to try this new way of getting the result for EFP.

$\quad$\\
{\it The case $n=5$}\\
First as above we write down consequences of the properties
(\ref{rec}) and (\ref{prop2}). The first one (\ref{rec}) is equivalent 
to the set of three relations which can be obtained from eqs.
(\ref{rec1}) by taking $k=2$
$$
A_{5,2}(z_1,z_2|z_3,z_4|\infty)\;=\;{1\over 2}\,A_{4,2}(z_1,z_2|z_3,z_4)
$$
$$
A_{5,1}(z_1,z_2|z_3,z_4,\infty)\,-\,{1\over 2} \biggl\{
A_{5,2}(z_1,z_2|\infty,z_3|z_4)\,+\,A_{5,2}(z_1,z_2|\infty,z_4|z_3)\biggr\}
\;=\;{1\over 2} A_{4,1}(z_1,z_2|z_3,z_4)
$$
$$
A_{5,0}\,-\,
{1\over 2}\biggl\{
A_{5,1}(\infty,z_1|z_2,z_3,z_4)\,+\,
A_{5,1}(\infty,z_2|z_1,z_3,z_4)\,+\,
$$
\be
+\,A_{5,1}(\infty,z_3|z_1,z_2,z_4)\,+\,
A_{5,1}(\infty,z_4|z_1,z_2,z_3)\biggr\}
\;=\;{1\over 2} A_{4,0}
\label{eq5inf}
\ee

The Property 2 (\ref{prop2}) for $n=5$ is equivalent to the following
set of five equations
$$
A_{5,2}(z_1,z_2|z_3,z_4|z_4+i)\,+\,\g(z_{34})A_{5,2}(z_1,z_2|z_3,z_4+i|z_4)
\;=\;0
$$
$$
\g(z_{14})A_{5,2}(z_1,z_4+i|z_2,z_4|z_3)\,+\,
\g(z_{24})A_{5,2}(z_1,z_4|z_2,z_4+i|z_3)
\;=\;0
$$

$$
\a(z_{34})A_{5,2}(z_1,z_2|z_3,z_4+i|z_4)\,-\,2\,A_{5,2}(z_1,z_2|z_4+i,z_4|z_3)
\,+\,A_{5,1}(z_1,z_2|z_3,z_4,z_4+i)\;=\;0
$$

$$
\a(z_{24})A_{5,2}(z_1,z_4|z_2,z_4+i|z_3)\,+\,
\a(z_{34})A_{5,2}(z_1,z_4|z_4+i,z_3|z_2)
\,+\,A_{5,1}(z_1,z_4|z_2,z_3,z_4+i)\,+\,
$$
$$
\g(z_{14})A_{5,1}(z_1,z_4+i|z_2,z_3,z_4)\;=\;0
$$

$$
A_{5,0}\,+\,\a(z_{14})A_{5,1}(z_1,z_4+i|z_2,z_3,z_4)\,+\,
\a(z_{24})A_{5,1}(z_2,z_4+i|z_1,z_3,z_4)\,+
$$
\be
\,+\,\a(z_{34})A_{5,1}(z_3,z_4+i|z_1,z_2,z_4)
\,-\,2\,A_{5,1}(z_4,z_4+i|z_1,z_2,z_3)\;=\;0
\label{eq5i}
\ee

The difference with above cases $n=2,3,4$ is that now up to the moment
we do not have a solution to these equations. Therefore we have to
get it. Let us briefly describe how we do this. First we take 
the ansatz (\ref{An0}-\ref{Q}), namely,
\be
A_{5,0}\;=\;{1\over 6}
\label{A50}
\ee
\be
A_{5,1}(z_1,z_2|z_3,z_4,z_5)\;=\;{{\ov Q}_{5,1}(z_1,z_2|z_3,z_4,z_5)\over
z_{13}z_{14}z_{15}z_{23}z_{24}z_{25}}
\label{A51}
\ee
\be
A_{5,2}(z_1,z_2|z_3,z_4|z_5)\;=\;{{\ov Q}_{5,2}(z_1,z_2|z_3,z_4|z_5)\over
z_{13}z_{14}z_{15}z_{23}z_{24}z_{25}z_{35}z_{45}}
\label{A52}
\ee
and demand that the polynomials ${\ov Q}_{5,1}(z_1,z_2|z_3,z_4,z_5)$
and ${\ov Q}_{5,2}(z_1,z_2|z_3,z_4|z_5)$ have the same symmetry properties
as the functions $A_{5,1}$ and $A_{5,2}$ and 
the same maximum powers of the variables $z_1,\ldots, z_5$ as their
denominators in the r.h.s. of (\ref{A51}) and (\ref{A52}) respectively. 
Namely, we write
\be
{\ov Q}_{5,1}(z_1,z_2|z_3,z_4,z_5)\;=\;\sum_{
{\small 
\begin{array}{c}
 0\le i_1,i_2\le 3, \quad 0\le i_3,i_4,i_5\le 2\\
i_1+\ldots + i_5\le 6 
\end{array}
}}
 C_{5,1}(i_1,i_2|i_3,i_4,i_5)\; z_1^{i_1}z_2^{i_2}z_3^{i_3}z_4^{i_4}z_5^{i_5}
\label{Q51ansatz}
\ee
\be
{\ov Q}_{5,2}(z_1,z_2|z_3,z_4|z_5)\;=\;\sum_{
{\small
\begin{array}{c}
0\le i_1,i_2,i_3,i_4\le 3, \quad 0\le i_5\le 4\\
i_1+\ldots + i_5\le 8 
\end{array}
}}
 C_{5,2}(i_1,i_2|i_3,i_4|i_5)\; z_1^{i_1}z_2^{i_2}z_3^{i_3}z_4^{i_4}z_5^{i_5}
\label{Q52ansatz}
\ee
with the coefficients $C_{5,1}(i_1,i_2|i_3,i_4,i_5)$ and 
$C_{5,2}(i_1,i_2|i_3,i_4|i_5)$
which have evident symmetry properties 
\be
C_{5,1}(i_1,i_2|i_3,i_4,i_5)=C_{5,1}(i_2,i_1|i_3,i_4,i_5)=
C_{5,1}(i_1,i_2|\s(i_3),\s(i_4),\s(i_5))
\label{C51}
\ee
where $\s$ is any element of the permutation group of three elements $S_3$
while
\be
C_{5,2}(i_1,i_2|i_3,i_4|i_5)=C_{5,2}(i_2,i_1|i_3,i_4|i_5)=
C_{5,2}(i_1,i_2|i_4,i_3|i_5)=
C_{5,2}(i_3,i_4|i_1,i_2|i_5)
\label{C52}
\ee

In accordance with eq. (\ref{transl}) we also demand the translational
invariance of these polynomials
\be
{\ov Q}_{5,1}(z_1+a,z_2+a|z_3+a,z_4+a,z_5+a)\;=\;
{\ov Q}_{5,1}(z_1,z_2|z_3,z_4,z_5)
\label{Q51transl}
\ee
\be
{\ov Q}_{5,2}(z_1+a,z_2+a|z_3+a,z_4+a|z_5+a)\;=\;
{\ov Q}_{5,2}(z_1,z_2|z_3,z_4|z_5)
\label{Q52transl}
\ee
These conditions allow us to fix some of the coefficients $C_{5,1}$ and
$C_{5,2}$ in (\ref{Q51ansatz}-\ref{Q52ansatz}).

The rest of these coefficients may be fixed by satisfying the
equations (\ref{eq5i}). One should say that in spite of
the fact that it turns out to be an over determined system it has 
a solution and this solution is {\it unique}. 
Also it is interesting to note that
after the equations (\ref{eq5i}) are satisfied 
all the residual equations (\ref{eq5inf}) are satisfied
automatically (!). 

Let us show the result
$$
{\ov Q}_{5,2}(z_1,z_2|z_3,z_4|z_5)\;=\;
$$
\be
{1\over 2}(z_{15}z_{25}-1)(z_{35}z_{45}-1)\,{\ov Q}_{4,2}(z_1,z_2|z_3,z_4)\,+\,
{1\over 360} (z_{12}^2+4)(z_{34}^2+4)(5-z_{13}z_{24}-z_{14}z_{23})
\label{Q52res}
\ee
$$
{\ov Q}_{5,1}(z_1,z_2|z_3,z_4,z_5)\;=\;
$$
$$
{1\over 2}(z_{15}z_{25}-1)\,{\ov Q}_{4,2}(z_1,z_2|z_3,z_4)\,+\,
{1\over 2}(z_{14}z_{24}-1)\,{\ov Q}_{4,2}(z_1,z_2|z_3,z_5)\,+\,
{1\over 2}(z_{13}z_{23}-1)\,{\ov Q}_{4,2}(z_1,z_2|z_4,z_5)
$$
\be
-\,{(z_{12}^2+4)\over 360}\biggl\{
22\,+\,(z_{34}^2-2)(z_{15}z_{25}+{1\over 2})
\,+\,(z_{35}^2-2)(z_{14}z_{24}+{1\over 2})
\,+\,(z_{45}^2-2)(z_{13}z_{23}+{1\over 2})\biggr\}
\label{Q51res}
\ee

Now using eqs. (\ref{A50}-\ref{A52}) and
substituting these formulae into the ansatz (\ref{Pninhom}) 
we get the answer for $P_5$ in the inhomogeneous case.


$\quad$\\
{\it The case $n=6$}\\
We proceed in the same line as for the case $n=5$.
The Property 1 i.e. eq.(\ref{rec}) or equivalently the set of the 
recurrent relations (\ref{rec2}) at $k=3$ gives us three equations 
$$
A_{6,2}(z_1,z_2|z_3,z_4|z_5,\infty)\,
-\,{1\over 2}A_{6,3}(z_1,z_2|z_3,z_4|z_5,\infty)
\;=\;{1\over 2}A_{5,2}(z_1,z_2|z_3,z_4|z_5)
$$
$$
A_{6,1}(z_1,z_2|z_3,z_4,z_5,\infty)\,-\,
{1\over 2} \biggl\{
A_{6,2}(z_1,z_2|\infty,z_3|z_4,z_5)\,+\,
A_{6,2}(z_1,z_2|\infty,z_4|z_3,z_5)\,+\,
$$
$$
+\,A_{6,2}(z_1,z_2|\infty,z_5|z_3,z_4)\biggr\}
\;=\;{1\over 2} A_{5,1}(z_1,z_2|z_3,z_4,z_5)
$$
$$
A_{6,0}\,-\,
{1\over 2}\biggl\{
A_{6,1}(\infty,z_1|z_2,z_3,z_4,z_5)\,+\,
A_{6,1}(\infty,z_2|z_1,z_3,z_4,z_5)\,+\,
A_{6,1}(\infty,z_3|z_1,z_2,z_4,z_5)\,+\,
$$
\be
+\, A_{6,1}(\infty,z_4|z_1,z_2,z_3,z_5)\,+\,
A_{6,1}(\infty,z_5|z_1,z_2,z_3,z_4)\biggr\}
\;=\;{1\over 2} A_{50}
\label{eq6inf}
\ee

The Property 2 (\ref{prop2}) for $n=6$ produces seven equations
$$
A_{6,2}(z_1,z_2|z_3,z_4|z_5,z_5+i)\,-\,2\,A_{6,3}(z_1,z_2|z_3,z_4|z_5,z_5+i)
\;=\;0
$$

$$
\g(z_{45})A_{6,3}(z_1,z_2|z_3,z_5|z_4,z_5+i)\,+\,
\g(z_{35})A_{6,3}(z_1,z_2|z_3,z_5+i|z_4,z_5)
\;=\;0
$$

$$
A_{6,2}(z_1,z_2|z_3,z_5|z_4,z_5+i)\,+\,
\g(z_{35})A_{6,2}(z_1,z_2|z_3,z_5+i|z_4,z_5)\,+\,
$$
$$
+\,\a(z_{45})A_{6,3}(z_1,z_2|z_3,z_5|z_4,z_5+i)\;=\;0
$$

$$
\a(z_{35})A_{6,2}(z_1,z_2|z_3,z_5+i|z_4,z_5)\,+\,
\a(z_{45})A_{6,2}(z_1,z_2|z_4,z_5+i|z_3,z_5)\,-\,
$$
$$
-\,2\,A_{6,2}(z_1,z_2|z_5,z_5+i|z_3,z_4)
\,+\,
A_{6,1}(z_1,z_2|z_3,z_4,z_5,z_5+i)\;=\;0
$$

$$
\g(z_{25})A_{6,2}(z_1,z_5|z_2,z_5+i|z_3,z_4)\,+\,
\g(z_{15})A_{6,2}(z_1,z_5+i|z_2,z_5|z_3,z_4)
\;=\;0
$$

$$
A_{6,1}(z_1,z_5|z_2,z_3,z_4,z_5+i)\,+\,
\g(z_{15})A_{6,1}(z_1,z_5+i|z_2,z_3,z_4,z_5)\,+\,
$$
$$
\a(z_{25})A_{6,2}(z_1,z_5|z_2,z_5+i|z_3,z_4)\,+\,
\a(z_{35})A_{6,2}(z_1,z_5|z_3,z_5+i|z_2,z_4)\,+\,
$$
$$
+\,\a(z_{45})A_{6,2}(z_1,z_5|z_4,z_5+i|z_2,z_3)
\;=\;0
$$

$$
A_{6,0}\,-\,2\,A_{6,1}(z_5,z_5+i|z_1,z_2,z_3,z_4)\,+\,
\a(z_{15})A_{6,1}(z_1,z_5+i|z_2,z_3,z_4,z_5)\,+\,
$$
$$
\a(z_{25})A_{6,1}(z_2,z_5+i|z_1,z_3,z_4,z_5)\,+\,
\a(z_{35})A_{6,1}(z_3,z_5+i|z_1,z_2,z_4,z_5)\,+\,
$$
\be
\,+\,\a(z_{45})A_{6,1}(z_4,z_5+i|z_1,z_2,z_3,z_4)
\;=\;0
\label{eq6i}
\ee

In the same line as we did it above for the case $n=5$ we can
solve the equations (\ref{eq6inf}) and 
(\ref{eq6i}) in framework of the ansatz 
(\ref{An0}-\ref{Q})
\be
A_{6,0}\;=\;{1\over 7}
\label{A60}
\ee
\be
A_{6,1}(z_1,z_2|z_3,z_4,z_5,z_6)\;=\;{{\ov Q}_{6,1}(z_1,z_2|z_3,z_4,z_5,z_6)
\over
z_{13}z_{14}z_{15}z_{16}z_{23}z_{24}z_{25}z_{26}}
\label{A61}
\ee
\be
A_{6,2}(z_1,z_2|z_3,z_4|z_5,z_6)\;=\;{{\ov Q}_{6,2}(z_1,z_2|z_3,z_4|z_5,z_6)
\over
z_{13}z_{14}z_{15}z_{16}z_{23}z_{24}z_{25}z_{26}z_{35}z_{36}z_{45}z_{46}}
\label{A62}
\ee
\be
A_{6,3}(z_1,z_2|z_3,z_4|z_5,z_6)\;=\;{{\ov Q}_{6,3}(z_1,z_2|z_3,z_4|z_5,z_6)
\over
z_{13}z_{14}z_{15}z_{16}z_{23}z_{24}z_{25}z_{26}z_{35}z_{36}z_{45}z_{46}}
\label{A63}
\ee
As above
we demand that the polynomials ${\ov Q}_{6,1}(z_1,z_2|z_3,z_4,z_5,z_6)$,
${\ov Q}_{6,2}(z_1,z_2|z_3,z_4|z_5,z_6)$
and \\
${\ov Q}_{6,3}(z_1,z_2|z_3,z_4|z_5,z_6)$ have the same symmetry properties
as the functions $A_{6,1}$, $A_{6,2}$ , $A_{6,3}$ respectively.
Analogous to eqs. (\ref{Q51ansatz}-\ref{Q52ansatz}) we write
\be
{\ov Q}_{6,1}(z_1,z_2|z_3,z_4,z_5,z_6)=\sum_{
{\small
\begin{array}{c}
0\le i_1,i_2\le 4,\quad 0\le i_3,i_4,i_5,i_6\le 2\\
i_1+\ldots + i_6\le 8 
\end{array}
}}
C_{6,1}(i_1,i_2|i_3,i_4,i_5,i_6)\; 
z_1^{i_1}z_2^{i_2}z_3^{i_3}z_4^{i_4}z_5^{i_5}z_6^{i_6}
\label{Q61ansatz}
\ee
\be
{\ov Q}_{6,2}(z_1,z_2|z_3,z_4|z_5,z_6)=\sum_{
{\small
\begin{array}{c}
0\le i_1,i_2,\ldots,i_6\le 4\\
i_1+\ldots + i_6\le 12 
\end{array}
}}
 C_{6,2}(i_1,i_2|i_3,i_4|i_5,i_6)\; 
z_1^{i_1}z_2^{i_2}z_3^{i_3}z_4^{i_4}z_5^{i_5}z_6^{i_6}
\label{Q62ansatz}
\ee
\be
{\ov Q}_{6,3}(z_1,z_2|z_3,z_4|z_5,z_6)\;=\;\sum_{
{\small
\begin{array}{c}
0\le i_1,i_2,\ldots,i_6\le 4\\
i_1+\ldots + i_6\le 12 
\end{array}
}}
 C_{6,3}(i_1,i_2|i_3,i_4|i_5,i_6)\; 
z_1^{i_1}z_2^{i_2}z_3^{i_3}z_4^{i_4}z_5^{i_5}z_6^{i_6}
\label{Q63ansatz}
\ee
with the coefficients $C_{6,1}(i_1,i_2|i_3,i_4,i_5,i_6)$,
$C_{6,2}(i_1,i_2|i_3,i_4|i_5,i_6)$ and $C_{6,3}(i_1,i_2|i_3,i_4|i_5,i_6)$
which satisfy the symmetry conditions
\be
C_{6,1}(i_1,i_2|i_3,i_4,i_5,i_6)=C_{6,1}(i_2,i_1|i_3,i_4,i_5,i_6)=
C_{6,1}(i_1,i_2|\s(i_3),\s(i_4),\s(i_5),\s(i_6))
\label{C61}
\ee
where $\s$ is any element of the permutation group of four elements $S_4$
$$
C_{6,2}(i_1,i_2|i_3,i_4|i_5,i_6)=C_{6,2}(i_2,i_1|i_3,i_4|i_5,i_6)=
C_{6,2}(i_1,i_2|i_4,i_3|i_5,i_6)=C_{6,2}(i_1,i_2|i_3,i_4|i_6,i_5)=
$$
\be
=C_{6,2}(i_3,i_4|i_1,i_2|i_5,i_6)
\label{C62}
\ee
$$
C_{6,3}(i_1,i_2|i_3,i_4|i_5,i_6)=C_{6,3}(i_2,i_1|i_3,i_4|i_5,i_6)=
C_{6,3}(i_1,i_2|i_4,i_3|i_5,i_6)=C_{6,3}(i_1,i_2|i_3,i_4|i_6,i_5)=
$$
\be
=C_{6,3}(i_3,i_4|i_1,i_2|i_5,i_6)=C_{6,3}(i_1,i_2|i_5,i_6|i_3,i_4)
\label{C63}
\ee

As in the previous case the translational invariance of these polynomials
\be
{\ov Q}_{6,1}(z_1+a,z_2+a|z_3+a,z_4+a,z_5+a,z_6+a)\;=\;
{\ov Q}_{6,1}(z_1,z_2|z_3,z_4,z_5,z_6)
\label{Q61transl}
\ee
\be
{\ov Q}_{6,2}(z_1+a,z_2+a|z_3+a,z_4+a|z_5+a,z_6+a)\;=\;
{\ov Q}_{6,2}(z_1,z_2|z_3,z_4|z_5,z_6)
\label{Q62transl}
\ee
\be
{\ov Q}_{6,3}(z_1+a,z_2+a|z_3+a,z_4+a|z_5+a,z_6+a)\;=\;
{\ov Q}_{6,3}(z_1,z_2|z_3,z_4|z_5,z_6)
\label{Q63transl}
\ee
allow us to fix a lot of the coefficients $C_{6,1}, C_{6,2}$ and
$C_{6,3}$.

Then with the help of computer 
we solve the over determined system of the equations 
(\ref{eq6i}). It is left to check that all
other equations (\ref{eq6inf}) are satisfied
automatically as in the case $n=5$.

The result looks as follows
$$
{\ov Q}_{6,3}(z_1,z_2|z_3,z_4|z_5,z_6)\;=\;
$$
\be
6^3\,{\ov Q}_{4,2}(z_1,z_2|z_3,z_4){\ov Q}_{4,2}(z_1,z_2|z_5,z_6)
{\ov Q}_{4,2}(z_3,z_4|z_5,z_6)\;+\;
(z_{12}^2+4)(z_{34}^2+4)(z_{56}^2+4)\,\Lambda_{6,3}(z_1,z_2|z_3,z_4|z_5,z_6)
\label{Q63res}
\ee
$$
{\ov Q}_{6,2}(z_1,z_2|z_3,z_4|z_5,z_6)\;=\;
$$
\be
{6^3\over 2}\,{\ov Q}_{4,2}(z_1,z_2|z_3,z_4){\ov Q}_{4,1}(z_1,z_2|z_5,z_6)
{\ov Q}_{4,1}(z_3,z_4|z_5,z_6)\;+\;
(z_{12}^2+4)(z_{34}^2+4)\,\Lambda_{6,2}(z_1,z_2|z_3,z_4|z_5,z_6)
\label{Q62res}
\ee
$$
{\ov Q}_{6,1}(z_1,z_2|z_3,z_4,z_5,z_6)\;=\;
$$
\be
{1\over 30}\,(z_{13}z_{23}-1)(z_{14}z_{24}-1)(z_{15}z_{25}-1)(z_{16}z_{26}-1)
\;+\;(z_{12}^2+4)\,\Lambda_{6,1}(z_1,z_2|z_3,z_4,z_5,z_6)
\label{Q61res}
\ee
where
$$
\Lambda_{6,3}(z_1,z_2|z_3,z_4|z_5,z_6)\;=\;
{1\over 5400}(5-z_{13}z_{24}-z_{14}z_{23})
(5-z_{15}z_{26}-z_{16}z_{25})(5-z_{35}z_{46}-z_{36}z_{45})\,
$$
\be
-\,
{1\over 189000}(z_{12}^2-26)(z_{34}^2-26)(z_{56}^2-26)\,+\,
{1\over 200} (z_{12}^2+z_{34}^2+z_{56}^2-16)
\label{lambda63}
\ee
\vspace{0.5cm}
$$
\Lambda_{6,2}(z_1,z_2|z_3,z_4|z_5,z_6)\;=\;
\biggl\{{1\over 2}\Lambda_{6,3}(z_1,z_2|z_3,z_4|z_5,z_6)\,+\,
{1\over 100}{\ov Q}_{42}(z_1,z_2|z_3,z_4)(z_{56}^2+1)\biggr\}(z_{56}^2+13)\,+
$$
$$
\biggl\{{1\over 720}(5-z_{13}z_{24}-z_{14}z_{23})
(z_{56}^2-3+z_{15}z_{26}+z_{16}z_{25}+z_{35}z_{46}+z_{36}z_{45})\,-\,
$$
\be
-\,{1\over 2520}(z_{12}^2+{11\over 2})(z_{34}^2+{11\over 2})\,-\,{1\over 160}
\biggr\}(z_{56}^2+1)
\label{lambda62}
\ee
\vspace{0.5cm}
$$
\Lambda_{6,1}(z_1,z_2|z_3,z_4,z_5,z_6)\;=\;
{1\over 600}(z_{13}z_{14}z_{23}z_{24}z_{56}^2\,+\,
z_{13}z_{15}z_{23}z_{25}z_{46}^2\,+\,\ldots
\,+\,z_{15}z_{16}z_{25}z_{26}z_{34}^2)\,+\,
$$
$$
+\,{11\over 900}(z_{13}z_{14}z_{23}z_{24}\,+\,
z_{13}z_{15}z_{23}z_{25}\,+\,\ldots
\,+\,z_{15}z_{16}z_{25}z_{26})\,-
$$
$$
\,-\,
{1\over 225}\biggl\{(z_{13}z_{24}+z_{14}z_{23})z_{56}^2\,+\,
(z_{13}z_{25}+z_{15}z_{23})z_{46}^2\,+\,\ldots
\,+\,(z_{15}z_{26}+z_{16}z_{25})z_{34}^2\biggr\}\,-
$$
$$
\,-\,
{7\over 300}\biggl\{(z_{13}z_{24}+z_{14}z_{23})\,+\,
(z_{13}z_{25}+z_{15}z_{23})\,+\,\ldots
\,+\,(z_{15}z_{26}+z_{16}z_{25})\biggr\}\,+\,
{1\over 8400}z_{12}^2(z_{34}^2z_{56}^2+z_{35}^2z_{46}^2+z_{36}^2z_{45}^2)\,-
$$
$$
\,-\,
{19\over 2520}(z_{34}^2z_{56}^2+z_{35}^2z_{46}^2+z_{36}^2z_{45}^2)\,+\,
{1\over 1575}z_{12}^2(z_{34}^2+z_{56}^2+z_{35}^2+z_{46}^2+z_{36}^2+z_{45}^2)\,-
$$
\be
\,-\,
{1\over 168}(z_{34}^2+z_{56}^2+z_{35}^2+z_{46}^2+z_{36}^2+z_{45}^2)\,+\,
{13\over 840}z_{12}^2\,+\,{29\over 70}
\label{Lambda61}
\ee

By means of this solution, eqs. (\ref{A60}-\ref{A63}) and
the ansatz (\ref{Pninhom}) for $n=6$
we come (for the very first time (!))
to the result for $P_6$ in the inhomogeneous case.

Let us make a remark about some additional amusing factorization property of
the polynomials ${\ov Q}_{nl}$ whose meaning is still to realize, namely
$$
{\ov Q}_{3,1}(z_1,z_1+2i|z_3)\;=\;{1\over 12}(z_{13}+i)^2
$$

$$
{\ov Q}_{4,2}(z_1,z_1+2i|z_3,z_4)\;=\;{1\over 36}(z_{13}+i)^2(z_{14}+i)^2
$$
$$
{\ov Q}_{4,1}(z_1,z_1+2i|z_3,z_4)\;=\;{1\over 18}(z_{13}+i)^2(z_{14}+i)^2
$$

$$
{\ov Q}_{5,2}(z_1,z_1+2i|z_3,z_4|z_5)\;=\;{1\over 6}
(z_{13}+i)^2(z_{14}+i)^2(z_{15}+i)^2\,{\ov Q}_{31}(z_3,z_4|z_5)
$$
$$
{\ov Q}_{5,1}(z_1,z_1+2i|z_3,z_4|z_5)\;=\;{1\over 24}
(z_{13}+i)^2(z_{14}+i)^2(z_{15}+i)^2
$$

$$
{\ov Q}_{6,3}(z_1,z_1+2i|z_3,z_4|z_5,z_6)\;=\;{1\over 6}
(z_{13}+i)^2(z_{14}+i)^2(z_{15}+i)^2(z_{16}+i)^2\,
{\ov Q}_{4,2}(z_3,z_4|z_5,z_6)
$$
$$
{\ov Q}_{6,2}(z_1,z_1+2i|z_3,z_4|z_5,z_6)\;=\;{1\over 6}
(z_{13}+i)^2(z_{14}+i)^2(z_{15}+i)^2(z_{16}+i)^2\,
{\ov Q}_{4,1}(z_3,z_4|z_5,z_6)
$$
$$
{\ov Q}_{6,1}(z_1,z_1+2i|z_3,z_4|z_5,z_6)\;=\;{1\over 30}
(z_{13}+i)^2(z_{14}+i)^2(z_{15}+i)^2(z_{16}+i)^2
$$
Looking at these formulae we can assume that
\be
{{\ov Q}_{n,l}(z_1,z_1+2i|z_3,z_4|\ldots|z_{2l-1}z_{2l}|z_{2l+1}\ldots z_n)
\over
{\ov Q}_{n-2,l-1}(z_3,z_4|\ldots|z_{2l-1}z_{2l}|z_{2l+1}\ldots z_n)}\;=\;
{1\over 6}\prod_{j=3}^n(z_{1j} + i)^2
\label{fact}
\ee

\section{The structure of $P(n)$ in homogeneous case}

As we have discussed in \cite{BKNS} the formulae in the homogeneous case 
become more regular in terms of the alternating zeta values 
\be
\zeta_a(s)\;=\;\sum_{n>0}{(-1)^{n-1}\over n^s}\;=\; - \mbox{Li}_s(-1)
\label{za}
\ee
where $\mbox{Li}_s(x)$ is the polylogarithm. 
The alternating zeta series is related to the Riemann zeta function 
as follows
\be
\zeta(s)\;=\;{1\over 1-2^{1-s}}\zeta_a(s)
\label{za1}
\ee

Unlike Riemann zeta function the alternating zeta series is regular at $s=1$, it is $\zeta_a(1)=\log 2$.
Let us mention that the answers for $P(1),\ldots, P(4)$ can be obtained
from the formulae (\ref{P1}-\ref{P4}) by taking the homogeneous limit i.e.
$z_j\rightarrow 0$ and using the expansion of the function $G$ (\ref{Gexp}).

The same can be also done for $n=5$ and $n=6$ by means of the formula (\ref{Pninhom}),
formulae (\ref{An0}), (\ref{Anl}), (\ref{Q}) and the answers for the polynomials
(\ref{Q52res}-\ref{Q51res}) and (\ref{Q63res}-\ref{Q61res}) for the cases
$n=5$ and $n=6$ respectively.

For  $n\le 5$ we reproduce the known results, see  \cite{BKNS}  formula (1.16) in  there.

In the case $n=6$ we discover a new  result 
\beq
&\ds P(6)\; =\; 
{1\over 7}
\biggl\{
1 - 35\,\zeta_a(1) + 322\,\zeta_a(3) - {9244\over 5}\,\zeta_a(5) 
 + {22694\over 5}\,\zeta_a(7) - 2982\,\zeta_a(9)
&\nonumber\\ 
&\ds      
 - {3920\over 3}\,\zeta_a(1)\cdot\zeta_a(3)
 + {369908\over 15}\,\zeta_a(1)\cdot\zeta_a(5)
 - {28784\over 5}\,\zeta_a(3)^2 
- {263816\over 3}\,\zeta_a(1)\cdot\zeta_a(7)
&\nonumber\\ 
&\ds      
 + {3458\over 15}\,\zeta_a(3)\cdot\zeta_a(5)
+{323344\over 5}\,\zeta_a(1)\cdot\zeta_a(9)
+ {933702\over 5}\,\zeta_a(3)\cdot\zeta_a(7)
 - {751592\over 9}\,\zeta_a(5)^2 
&\nonumber\\ 
&\ds      
 - {2627842\over 15}\,\zeta_a(3)\cdot\zeta_a(9) 
+{235963\over 9}\,\zeta_a(5)\cdot\zeta_a(7)
+{368564\over 3}\,\zeta_a(5)\cdot\zeta_a(9)
 - {644987\over 9}\,\zeta_a(7)^2
&\nonumber\\ 
&\ds      
+{538496\over 45}\,\zeta_a(1)\cdot\zeta_a(3)\cdot\zeta_a(5)
-{269248\over 135}\,\zeta_a(3)^3 
-{1143268\over 9}\,\zeta_a(1)\cdot\zeta_a(3)\cdot\zeta_a(7)
+ {653296\over 9}\,\zeta_a(1)\cdot\zeta_a(5)^2
&\nonumber\\ 
&\ds      
-{163324\over 45}\,\zeta_a(3)^2\cdot\zeta_a(5) 
+{1737148\over 15}\,\zeta_a(1)\cdot\zeta_a(3)\cdot\zeta_a(9)
-{1737148\over 45}\,\zeta_a(3)^2\cdot\zeta_a(7)
+{124082\over 9}\,\zeta_a(3)\cdot\zeta_a(5)^2
&\nonumber\\ 
&\ds      
 - {528164\over 3}\,\zeta_a(1)\cdot\zeta_a(5)\cdot\zeta_a(9)
+ {924287\over 9}\,\zeta_a(1)\cdot\zeta_a(7)^2 
+ {264082\over 5}\,\zeta_a(3)^2\cdot\zeta_a(9)
&\nonumber\\ 
&\ds      
-{264082\over 9}\,\zeta_a(3)\cdot\zeta_a(5)\cdot\zeta_a(7) 
+{188630\over 27}\,\zeta_a(5)^3 \biggr\}
&\label{P6}
\eeq

This support our hypothesis that all $P(n)$ can be expression in terms of values of Riemann zeta functions at odd
arguments, $\log 2$ and rational coefficients.

From this expression we can get numerical value

\be
P(6) = 7.068127533\cdot 10^{-9}
\label{number}
\ee
In \cite{BKNS} a  numerical
method was used for evaluation of $P(6)$. It is called Density Matrix Renormalization Group (DMRG).
The results can be found in 
Table 1 in \cite{BKNS}. In particular 
$P(6)=7.05\cdot 10^{-9}$ with an uncertainty in the second digit after the  decimal
 point. It is in a good agreement with our analytic
result (\ref{P6}), (\ref{number}).

Looking at the formulae (1.16) of \cite{BKNS} and the above expression (\ref{P6}) 
we can  make a general conjecture for the dependence of $P(n)$
on the alternating zeta series
\be
P(n)\;=\;{1\over (n+1)}\sum_{{\vec r}\in U} B^{(n)}_{r_0,r_1,\ldots,r_{n-2}}
\prod_{j=0}^{n-2} {[\zeta_a(2\,j+1)]}^{r_j}
\label{Pn}
\ee
where all coefficients $B^{(n)}_{r_0,r_1,\ldots,r_{n-2}}$ are rational 
and the sum is over non-negative integers $r_0,\ldots,r_{n-2}$ which
belong to the region $U$ determined by the following two conditions
\beq
&\ds \sum_{j=0}^{n-2} r_j\;\le [n/2] & \nonumber\\
&\ds \sum_{j=0}^{n-2} r_j\,(2j+1)\;\le \frac{n(n-1)}{2}&
\label{U}
\eeq

Let us show how the non-zero coefficients 
$B^{(n)}_{r_0,r_1,\ldots,r_{n-2}}$ look like for 
the cases when we know the manifest analytic answer, 
namely, when $n=1,2,\ldots 6$
\beq
& B^{(1)}\;=\;1 & \nonumber\\
&\quad &\nonumber\\
& B^{(2)}_{0}\;=\;1\quad B^{(2)}_{1}\;=\;-1 & \nonumber\\
&\quad &\nonumber\\
& B^{(3)}_{0,0}\;=\;1\quad B^{(3)}_{1,0}\;=\;-4 
\quad B^{(3)}_{0,1}\;=\;2 & \nonumber\\
&\quad &\nonumber\\
&\ds  B^{(4)}_{0,0,0}\;=\;1
\quad B^{(4)}_{1,0,0}\;=\;-10 
\quad B^{(4)}_{0,1,0}\;=\;\frac{173}{9} 
\quad B^{(4)}_{0,0,1}\;=\;-\frac{110}{9}
& \nonumber\\
&\ds  B^{(4)}_{1,1,0}\;=\;-\frac{110}{9}
\quad B^{(4)}_{1,0,1}\;=\;\frac{170}{9} 
\quad B^{(4)}_{0,2,0}\;=\;-\frac{17}{3} & \nonumber\\
&\quad &\nonumber\\
&\ds  B^{(5)}_{0,0,0,0}\;=\;1
\quad B^{(5)}_{1,0,0,0}\;=\;-20 
\quad B^{(5)}_{0,1,0,0}\;=\;\frac{281}{3} 
\quad B^{(5)}_{0,0,1,0}\;=\;-\frac{1355}{6}
\quad B^{(5)}_{0,0,0,1}\;=\;\frac{889}{6}
& \nonumber\\
&\ds  B^{(5)}_{1,1,0,0}\;=\;-180 
\quad B^{(5)}_{1,0,1,0}\;=\;\frac{3920}{3} 
\quad B^{(5)}_{1,0,0,1}\;=\;-\frac{3290}{3} 
\quad B^{(5)}_{0,1,1,0}\;=\;-\frac{170}{3} \quad B^{(5)}_{0,1,0,1}\;=\;679
& \nonumber\\
&\ds B^{(5)}_{0,2,0,0}\;=\;-326  \quad B^{(5)}_{0,0,2,0}\;=\;-\frac{970}{3} &\nonumber\\
&\quad &\nonumber\\
&\ds
B^{(6)}_{0, 0, 0, 0, 0} = 1
&\nonumber\\ 
&\ds      
B^{(6)}_{1, 0, 0, 0, 0} = -35\quad
B^{(6)}_{0, 1, 0, 0, 0} = 322\quad
B^{(6)}_{0, 0, 1, 0, 0} = -{9244\over 5}\quad
B^{(6)}_{0, 0, 0, 1, 0} =  {22694\over 5}\quad
B^{(6)}_{0, 0, 0, 0, 1} =  -2982\quad
&\nonumber\\ 
&\ds      
B^{(6)}_{1, 1, 0, 0, 0} = -{3920\over 3}\quad
B^{(6)}_{1, 0, 1, 0, 0} = {369908\over 15}\quad
B^{(6)}_{1, 0, 0, 1, 0} = -{263816\over 3}\quad
B^{(6)}_{1, 0, 0, 0, 1} = {323344\over 5}\quad
&\nonumber\\ 
&\ds      
B^{(6)}_{0, 1, 1, 0, 0} = {3458\over 15}\quad
B^{(6)}_{0, 1, 0, 1, 0} = {933702\over 5}\quad
B^{(6)}_{0, 1, 0, 0, 1} = -{2627842\over 15}\quad
B^{(6)}_{0, 0, 1, 1, 0} = {235963\over 9}\quad
&\nonumber\\ 
&\ds      
B^{(6)}_{0, 0, 1, 0, 1} = {368564\over 3}\quad
B^{(6)}_{0, 2, 0, 0, 0} = -{28784\over 5}\quad
B^{(6)}_{0, 0, 2, 0, 0} = -{751592\over 9}\quad
B^{(6)}_{0, 0, 0, 2, 0} = -{644987\over 9}\quad
&\nonumber\\ 
&\ds      
B^{(6)}_{1, 1, 1, 0, 0} = {538496\over 45} \quad
B^{(6)}_{1, 1, 0, 1, 0} =  -{1143268\over 9}\quad
B^{(6)}_{1, 1, 0, 0, 1} = {1737148\over 15}\quad
B^{(6)}_{1, 0, 1, 0, 1} = -{528164\over 3} \quad
&\nonumber\\ 
&\ds      
B^{(6)}_{0, 1, 1, 1, 0} = -{264082\over 9}\quad
B^{(6)}_{1, 0, 2, 0, 0} = {653296\over 9}\quad
B^{(6)}_{1, 0, 0, 2, 0} = {924287\over 9}\quad
B^{(6)}_{0, 1, 2, 0, 0} = {124082\over 9}\quad
&\nonumber\\ 
&\ds      
B^{(6)}_{0, 2, 1, 0, 0} =-{163324\over 45} \quad
B^{(6)}_{0, 2, 0, 1, 0} = -{1737148\over 45}\quad
B^{(6)}_{0, 2, 0, 0, 1} = {264082\over 5}\quad
&\nonumber\\ 
&\ds      
B^{(6)}_{0, 3, 0, 0, 0} = -{269248\over 135}\quad
B^{(6)}_{0, 0, 3, 0, 0} = {188630\over 27}\quad
&\label{B}
\eeq

For us it was a bit 
surprising that two of coefficients $B^{(6)}$ appeared to be zero, namely,
\be
B^{(6)}_{1, 0, 1, 1, 0} = 0 \quad 
B^{(6)}_{1, 2, 0, 0, 0} = 0
\label{B6zero}
\ee
It means that the structures $\zeta_a(1)\cdot\zeta_a(5)\cdot\zeta_a(7)$
and  $\zeta_a(1)\cdot\zeta_a(3)^2$ do not appear in the final answer
for $P(6)$. Meanwhile the term $\zeta_a(3)^3$ survived with the
non-zero coefficient $B^{(6)}_{0, 3, 0, 0, 0}$ which we expected to
be zero.

The most evident conjecture that we can make looking at the formulae (\ref{B}) 
is as follows
\be
B^{(n)}_{0,0,\ldots,0}\;=\;1, 
\label{B0}
\ee
\be
B^{(n)}_{1,0,\ldots,0}\;=\;-\,\left(\begin{array}{c} n+1\\3\end{array}\right)
\label{B1}
\ee
where $\left(\begin{array}{c} n\\m\end{array}\right)$ 
is the binomial coefficient.

The next our conjecture is less trivial. 
As appeared the coefficients $B$ might satisfy some equations. 
One of them has the following form
\be
\sum_{r_0+r_1+\ldots +r_{n-2}\,=\,p} 
B^{(n)}_{r_0,r_1,\ldots,r_{n-2}}\;=\;(-1)^p
\left(\begin{array}{c} n-p\\p\end{array}\right) 
\label{sumB}
\ee
where $p$ is some fixed positive integer. 
The expression (\ref{sumB}) can be easily verified for the first coefficients
given by the formulae (\ref{B}).
For example, when $p=1$ we get the following equation
\be
B^{(n)}_{1,0,\ldots,0}\,+\,B^{(n)}_{0,1,\ldots,0}\,+\ldots\,+\,
B^{(n)}_{0,0,\ldots,1}\;=\;-\,n\,+\,1
\label{sumB1}
\ee
where we have already made the conjecture (\ref{B1}) for the first term.
We believe that there should be more equations like (\ref{sumB}) 
which probably provide the rigorous expression for the
coefficients $B$.

We have pointed out in our previous work \cite{BKNS} that since 
$\zeta_a(1), \zeta_a(3), \zeta_a(5), \ldots $ are very likely different 
irrational (or even transcendental) numbers 
\footnote{It was proven by Ap{\'e}ry \cite{Ap} that $\zeta(3)$ is irrational. Then 
Rivoal \cite{Riv} proved that one of the nine numbers $\zeta(5),\ldots,\zeta(21)$ is irrational.
One of the most recent theorem proved by Zudilin \cite{Zud} says that one
of the four values $\zeta(5),\ldots,\zeta(11)$ is irrational. (See also the
paper by D. Zagier \cite{za} and the paper by Yu. Nesterenko \cite{Nest}). }
$P(n)$ seem to be 
different irrational (or transcendental) numbers as well. 
This means that $P(n)$ does not satisfy polynomial recursion relation
with respect to the distance.

In order to clarify the structure of the formula (\ref{Pn}) let us 
formally replace 
all $\zeta_a(2j+1)$ by one complex variable $x$. This will define a new
function $P(n,x)$.

Then we can calculate $P(n,x)$ using our conjectures (\ref{Pn}), 
(\ref{sumB}) and some properties of the binomial coefficients.
Namely, 
\be
P(n,x)\;=\;
\sum_{p=0}^{[n/2]}(-x)^p\left(\begin{array}{c} n-p\\p\end{array}\right)\;=\;
{A_+^{n+1}-A_-^{n+1}\over (n+1)\sqrt{1-4 x}}
\label{pnx}
\ee
where 
\be
A_{\pm}\;=\;{1\pm \sqrt{1- 4 x}\over 2}
\label{Apm}
\ee
In particular when $x\rightarrow 1$ one gets
\be
P(n,1)\;=\;\cases{
\;\; 1,  \quad n  =  6 k\quad \mbox{or} \quad n = 6 k+1 \cr
\;\; 0,  \quad n  =  3 k+2 \cr
-1,  \quad n  =  6 k+3\quad  \mbox{or} \quad 6 k+4 \cr
}
\label{Pn1}
\ee
Note that the  alternating zeta values 
approaches  $1$ as the argument gos to infinity 

\be
\lim_{s\rightarrow\infty}\zeta_a(s)\;=\;1
\label{limit}
\ee
Another nice form of the formula (\ref{pnx}) can be obtained by substitution
\be
x = {1\over {4\cosh^2 {\a}}}
\label{xal}
\ee

Then
\be
P(n,{1\over 4\cosh^2{\a}})={\sinh{[(n+1)\a]}\over 2^n(n+1)\cosh^n{\a}
\sinh{\a}}
\label{Pnal}
\ee
We see that when $\a$ tends to zero 
\be
\lim_{\a\rightarrow 0}P(n,{1\over 4\cosh^2{\a}})={1\over 2^n}
\label{Pnal0}
\ee
Let us remark that this result appeared to coincide with the limiting formula 
(\ref{inftylimit}). We do not know if this is  accidental or there is
 a reason for this.

Let us briefly discuss the generating function for the values $P(n)$
\be
\Psi(y)\;=\;\sum_{n=0}^{\infty}y^n \, P(n)
\label{gen1}
\ee
where $P(0)=1$ by definition.
Taking into account the conjectures (\ref{B0}) and (\ref{B1}) 
we can easily get the two first terms for
the generating function $\Psi(y)$
\be
\Psi(y)\;=\;-{\ln{(1-y)}\over y}\;+\;{y^2\over 3(1-y)^3}\ln{2}\;+\;\ldots
\label{Psi}
\ee
As we discussed in \cite{BK1} and \cite{BKNS} we expect that for
$n\gg 1$ 
\be
P(n)\sim e^{-\kappa n^2}
\label{assPn}
\ee
If we substitute it formally into eq. (\ref{gen1}) with $y=e^{u}$ then
we can expect that
\be
\Psi(e^u)\sim \tilde\Psi(u)\;=\;\sum_{n=0}^{\infty} e^{-\kappa n^2 + u\,n}
\label{psiu}
\ee
so that the function $\tilde\Psi(u)$ satisfies the functional equation
\be
\tilde\Psi(u)\,+\,\tilde\Psi(-u)\,-\,1\;=\;\theta_3({iu\over 2},e^{-\kappa})
\label{funrel}
\ee
where $\theta_3$ is the third Jacobi 
theta function with the nome $e^{-\kappa}$.
So we can expect that the generating function (\ref{gen1}) may be related
with the elliptic functions and have some automorphic properties.

Another possibility is to put into the r.h.s. of eq. (\ref{gen1}) 
the values $P(n,x)$ given by (\ref{pnx}) instead of $P(n)$ then
\be
\Psi(x,y)\;=\;\sum_{n=0}^{\infty}y^n \, P(n,x)\; =\;
{1\over y\sqrt{1-4x}}\ln{2-y+y\sqrt{1-4x}\over 2-y-y\sqrt{1-4x}}
\label{gen1x}
\ee
Another generating function looks much simpler
\be
\Psi'(x,y)\;=\;\sum_{n=0}^{\infty}(n+1)y^n \, P(n,x)\; =\;
{1\over 1\,-\,y\,+\,x y^2}
\label{gen1x'}
\ee

\section{Discussion}

The main point of this communication was to consider the emptiness
formation probability in the inhomogeneous case. The basic advantage
of the inhomogeneous case is that we have more parameters at our
disposal and relation to the qKZ. 
We derived three general properties of $P_n$ from qKZ,
which appeared to be extremely useful.
From our experience of the direct calculation of $P_n$ with $n\le 4$
we conjectured a general ansatz for $P_n$ (\ref{Pninhom}).
 We   established that the ansatz (\ref{Pninhom}) with the first and
the second property from the Section 2 [ namely (\ref{vanish})]   
 completely fix the answer. The third property 
(\ref{norm}) turns out to be a corollary. This observation allowed
us to evaluate $P_5$ and $P_6$ very efficiently [ in the
inhomogeneous case]. The homogeneous limit of $P_5$ reproduced
the expression, which we obtained   in paper \cite{BKNS}  by going
through  hard and long computations. That time we were not sure
that we will be able to calculate $P_6$ at all.
Now it is  possible to
do it very quickly just by taking the homogeneous
limit of our result in the inhomogeneous case.
In the next publications we are planning to prove our general ansatz (\ref{Pninhom})
and evaluate $P(n)$ for arbitrary $n$. 
Actually we think that qKZ approach is so powerful that we will be
able to evaluate any correlation function in the Heisenberg XXX model and 
to show that it has a  structure similar to $P(n)$
\be
F(z_1,\ldots,z_n)_{\e_1,\ldots,\e_n}\;=\;\sum_{l=0}^{[{n\over 2}]} 
\bigl\{
A_{n,l}^{\e_1,\ldots,\e_n}
\prod_{j=1}^lG(z_{2j-1}-z_{2j}) + \mbox{permutations}\bigr\}
\label{correl}
\ee
where $\e_j=\pm 1$
\be
F(z_1,\ldots,z_n)_{\e_1,\ldots,\e_n}\;=\; 
\langle {\rm GS} | \prod^n_{j=1} P_j^{\e_j} | {\rm GS} \rangle,
\ee
and 
$$
P_j^{\pm}={1\pm \s^z_j\over 2}
$$
We may also expect that similar to the functions 
$A_{n,l}=A_{n,l}^{+,\ldots,+}$ from (\ref{Pninhom}) the functions 
$A_{n,l}^{\e_1,\ldots,\e_n}$
are rational of their arguments $z_1,\ldots,z_n$ also.

\section{Acknowledgements}

The authors would like to thank A. Abanov, R. Flume,  M. Jimbo, D. Kreimer,   M. Lashkevich, S. Lukyanov,  T. Miwa,
 Yu. Nesterenko, P. Pyatov, V. Tarasov, D. Zagier and A. Zamolodchikov for useful discussions.
This research  has been supported by the following grants:
NSF grant PHY-9988566, the Russian Foundation of Basic Research
under grant \# 01--01--00201, by INTAS under grants \#00-00055 and \# 00-00561.
HEB would like to thank the administration of the Max-Planck Institute for Mathematics
for hospitality and perfect conditions for the work.
HEB and FAS are grateful to organizers of the International Workshop 
{\it ``Conformal Field Theory and Integrable Models''}, Chernogolovka, September 2002, Russia
for opportunity to present this work.

\end{document}